\documentclass{ws-procs10x7}
\usepackage{balance}

\def\ipb{{\rm pb}^{-1}}
\def\ifb{{\rm fb}^{-1}}

\newcolumntype{d}[1]{D{.}{.}{#1}}

\makeindex
\begin{document}

\title{Hadronic Charm Decays in CLEO}

\author{Steven R. Blusk}

\address{Department of Physics, Syracuse University, Syracuse, NY 13244\\
E-mail: sblusk@phy.syr.edu}


\twocolumn[\maketitle\abstract{We present several results on
hadronic $D$ meson decays. We report on results from the
scan of the energy region from 3970 MeV to 4260 MeV, which was
used to determine an optimal energy to carry out the $D_s$ physics
program of CLEO-c. Improved measurements of inclusive and
exclusive $D$ and $D_s$ branching fractions are presented. We
also show results on Dalitz analyses of $D^+\to\pi^+\pi^-\pi^+$
using CLEO-c data and $D^0\to K^+K^-\pi^0$ using CLEO III data.}
]

\section{Introduction}

    The study of leptonic and semileptonic charm decays provide 
direct access to CKM elements, decay constants and form factors. 
Measurements of hadronic branching fractions (BF) and their Dalitz structure
provide important inputs to $B$-physics measurements undertaken at B-factories. 
CLEO has collected 281$~\ipb$ of data at the $\psi(3770)$ and, 
as of this time, an additional $\sim350~\ipb$
at $\sqrt{s}=4170$ MeV. In this report, we describe several recent hadronic charm
analyses from CLEO-c, and one from CLEO-III ($\sqrt{s}\approx~10$~GeV).

\section{The $D_s$ Energy Scan}

CLEO scanned the energy region from 3970-4260 MeV with the primary goal of determining
an optimal energy for $D_s$ physics. 
The scan included 12 points with a total luminosity of about 60~$\ipb$.
Candidate $D^0$, $D^+$ and $D_s$ mesons are reconstructed in 3, 5, and 8
modes, respectively.  Because of the highly constrained kinematics, the 
momentum of the candidate can be used to determine the final state, e.g.,
$D\bar{D}$, $D^*\bar{D}$, $D^*\bar{D}^*$, $D_s\bar{D_s}$, or $D_s^*\bar{D}_s$.



\begin{figure}
  \centering
  \includegraphics[height=.27\textheight]{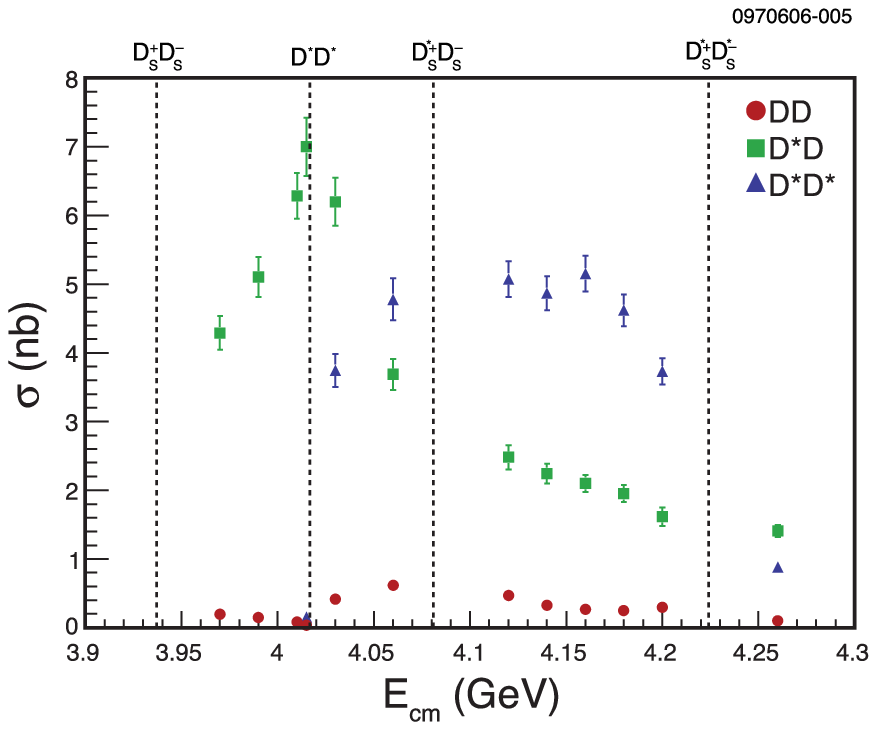}
  \includegraphics[height=.27\textheight]{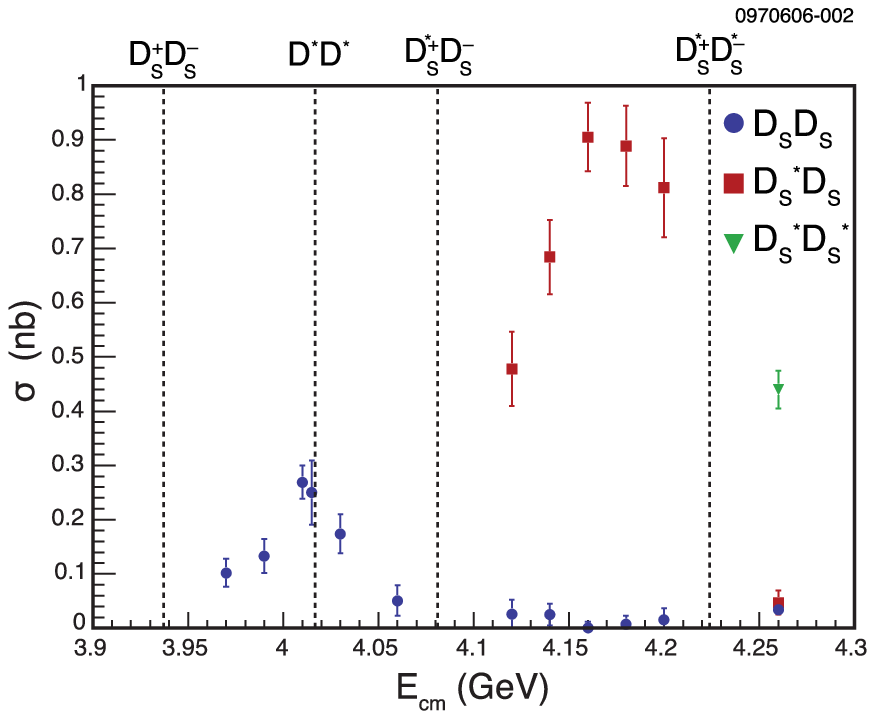}
  \caption{Measured production cross-section for $D\bar{D}$, $D\bar{D^*}$, and
$D^*\bar{D^*}$  (top) and
$D_s\bar{D}_s$, $D_s\overline{D_s^{*}}$, and $D_s^{*}\overline{D_s^{*}}$ (bottom)
l)
as a function of $\sqrt{s}$. The various $D_{(s)}$ pair thresholds are indicated.
\label{fig:ddbar}}
\end{figure}

We measure charm cross-sections in three ways: 
(1) inclusive hadronic event counting, with the non-charm components
subtracted, (2) inclusive $D_{(s)}$ counting, and (3) sum of 
$D_{(s)}^{(*)}\bar{D}_{(s)}^{(*)}$ exclusive final states.
We show in Fig.~\ref{fig:ddbar} the
$D\bar{D}$ (top) and $D_s\bar{D_s}$ (bottom) cross-sections using exclusive 
final states. A peak cross-section of about 0.9 nb for
$D_s^*\bar{D_s}$ is observed near 4170 MeV, which was then selected as the
optimal energy for the CLEO-c $D_s$ physics program.

\begin{figure}
  \centering
  \includegraphics[height=.28\textheight]{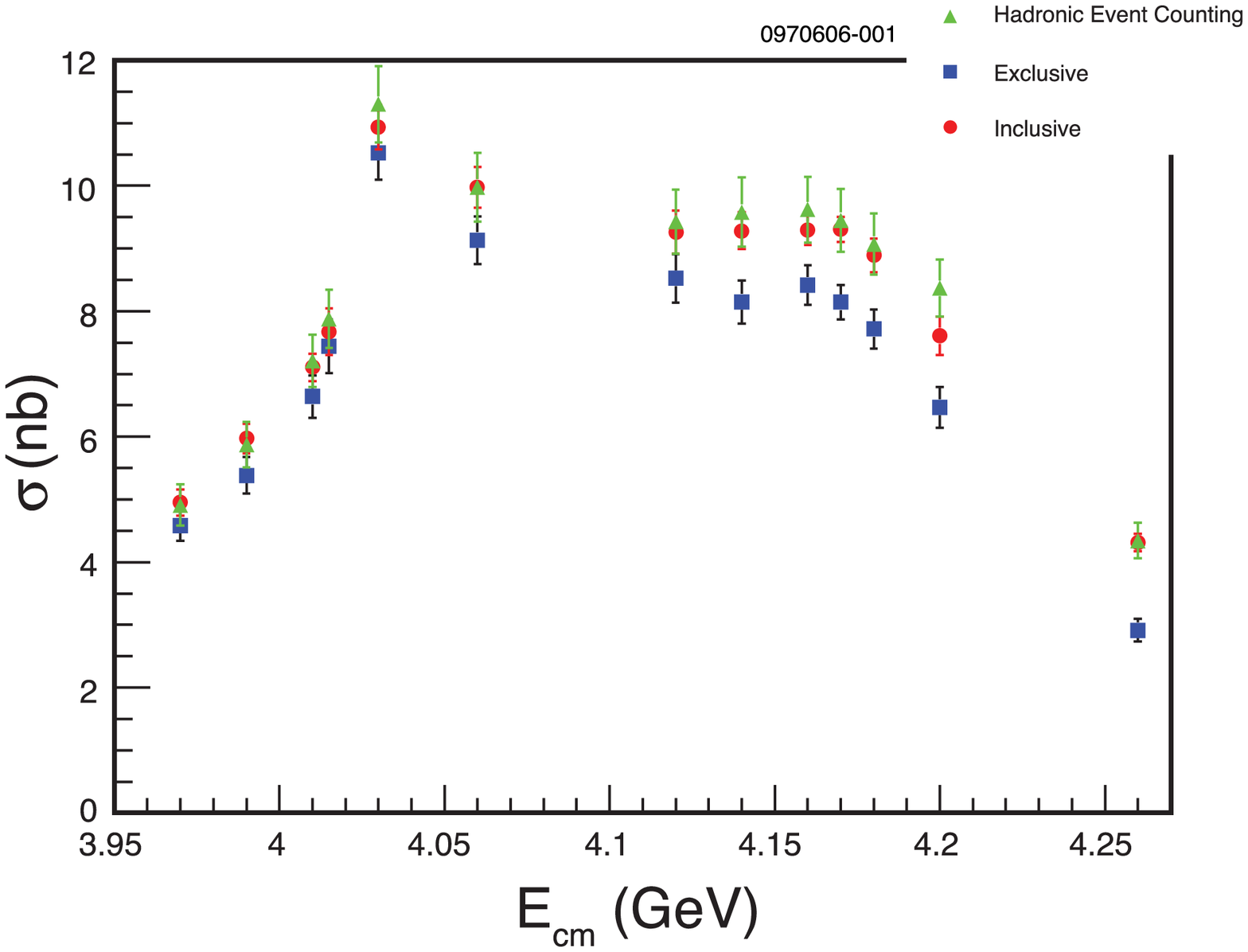}
  \includegraphics[height=.25\textheight]{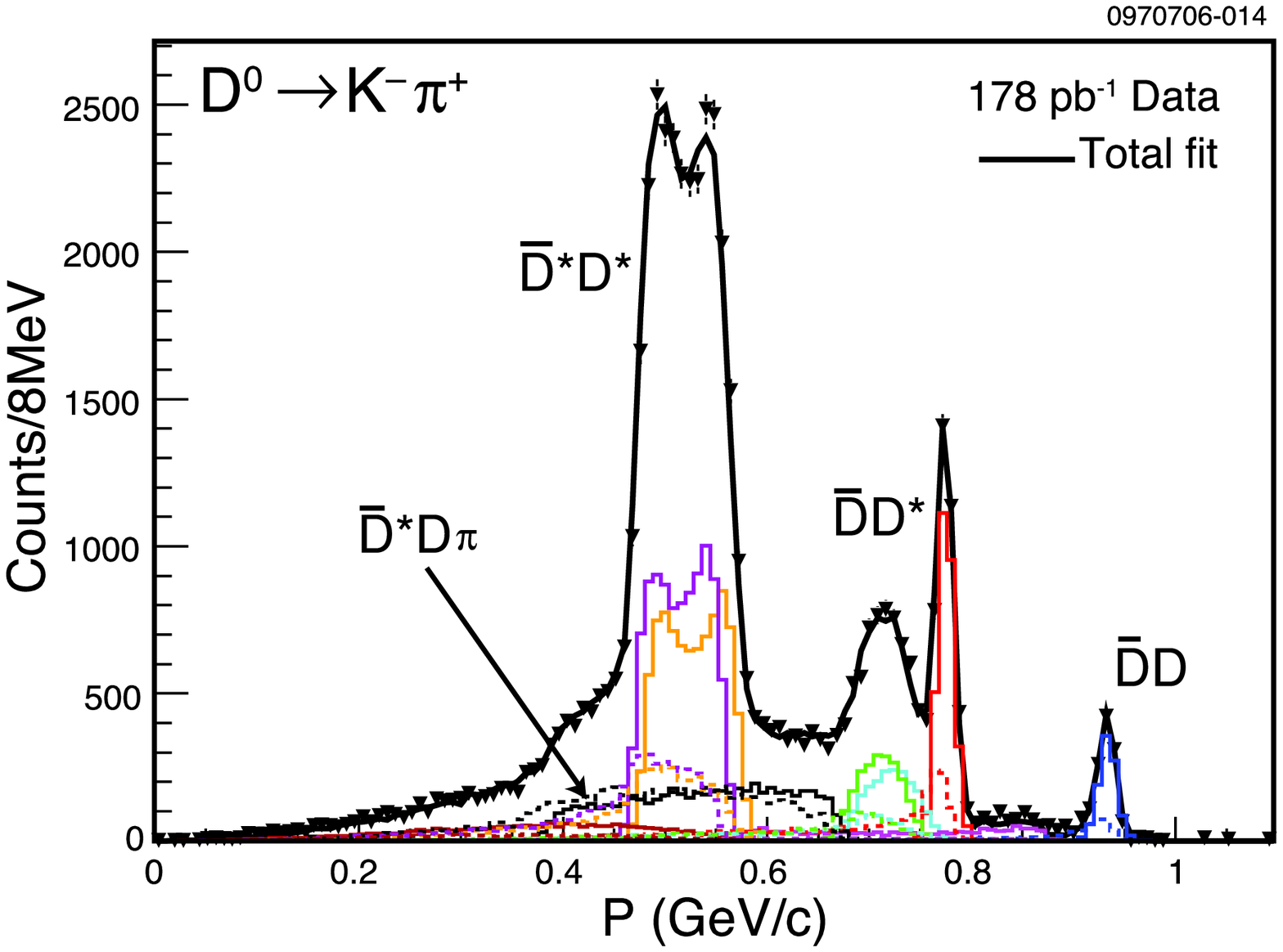}
  \caption{(Top) Measured charm cross-section using hadron counting, inclusive $D_{(s)}$ yields
and exclusive $D_{(s)}^{(*)}\bar{D}_{(s)}^{(*)}$ final states. (Bottom) Momentum spectrum
of $D^0$ mesons at $E_{\rm cm}=4160$~MeV in data (points) and simulation (lines). 
\label{fig:charm_xs}}
\end{figure}

The total charm cross-sections obtained using the three different methods are compared
in Fig.~\ref{fig:charm_xs} (top). The inclusive measurements show a clear excess over
exclusive reconstruction of $D_{(s)}^{(*)}\bar{D}_{(s)}^{(*)}$. 
Much of this excess can be attributed to a $D^*\bar{D}\pi$ final state. We model this
excess as a non-resonant contribution, and show in Fig.~\ref{fig:charm_xs} (bottom)
the momentum spectrum of $D^0\to K^-\pi^+$ decays using 178~$\ipb$ of data at 
$E_{\rm cm}=4170$~MeV, compared
to a simulation which includes the multi-body contribution. The 
prominent peaks from
two-body decays are indicated along with the broader spectrum from $\bar{D}^*D\pi$.
With the three-body final state included, a good description of the data is obtained.

\section{Hadronic Branching Fractions}

Production of $D\bar{D}$ and $D_s\bar{D_s}$ allows CLEO-c to determine absolute
BF's as well as $N_{D\bar{D}}$ using yields of double-tagged and
single-tagged events. Results based on 57~$\ipb$ have already
been published\cite{dhad}, yielding $D$ BF's in Cabibbo-favored
modes which are competitive with or better than the world average. Results based on 
281~$\ipb$ are imminent, and should provide uncertainties at the level of 1.5\% on the
key normalization modes: $D^0\to K^-\pi^+$ and $D^+\to K^-\pi^+\pi^+$. 

Measurements of Cabibbo-suppressed decays using 281~$\ipb$ have also been
published\cite{cs_prl}. Six new decay modes and eight improved measurements
were reported. Improved $D\to\pi\pi$ BF measurements
were used to perform an isospin analysis, yielding an amplitude ratio
$A_2/A_0=0.420\pm0.014\pm0.016$ and strong phase shift $\delta_I=(86.4\pm2.8\pm3.3)^o$
between the $\Delta I=3/2$ and $\Delta I=1/2$ isospin amplitudes.

We also report on preliminary measurements of BF's in $D_s$ decays
using $\sim$200~$\ipb$ of data collected at $E_{\rm cm}=4170$~MeV. 
As with the $D$ hadronic analysis\cite{dhad}, we use yields from single-tag modes
($D_s\to K_SK^+$, $K^+K^-\pi^+$, $K^+K^-\pi^+\pi^0$,$\pi^+\pi^-\pi^+$, 
$\eta\pi^+$, $\eta^{\prime}\pi^+$) and 36 double-tag modes to fit for the
absolute $D_s$ BF's. Table~\ref{tab:ds} shows the 
single-tag yields, average $D_s$ efficiencies 
and the BF's from the fit (double-tag yields not shown).
While these results are preliminary, the uncertainties are already significantly smaller 
than the 2006 world averages. The decay $D_s\to\phi\pi$ is a sub-mode of $K^+K^-\pi^+$,
and exraction of its branching fraction is complicated due to
interference with the nearby $f_0$. Alternately, we measure
the partial BF, consisting of events with $M_{K^+K^-}-M_{\phi}<10$~MeV.
The resulting partial BF
${\cal{B}}_{M_{K^+K^-}-M_{\phi}<10~{\rm MeV}}(D_s\to K^+K^-\pi)$ 
is $(1.98\pm0.12\pm0.09)\%$.


\begin{table}
\tbl{Preliminary results of $D_s$ branching fractions based on
$\sim$200~$\ipb$ of data collected at $E_{\rm cm}=4170$~MeV. Yields
are the sum of $D_s^+$ and $D_s^-$.
\label{tab:ds}}
{\begin{tabular}{@{}lccc@{}} 
\toprule
Mode                 &  Yield       & Eff.  &        BF \\
$D_s^{\pm}\to$       &              & (\%)  &      (\%)     \\
\hline
$K_SK^{\pm}$             & $1983\pm54$  & 37.5  & $1.50\pm0.09\pm0.05$ \\
$K^+K^-\pi^{\pm}$        & $8666\pm126$ & 44.3  & $5.57\pm0.30\pm0.19$ \\
$K^+K^-\pi^{\pm}\pi^0$   & $2410\pm119$ & 12.5  & $5.62\pm0.33\pm0.51$ \\
$\pi^{\pm}\pi^{\mp}\pi^{\pm}$    & $1916\pm112$ & 49.4  & $1.12\pm0.08\pm0.05$  \\
$\eta\pi^{\pm}$          & $1117\pm70$  & 19.5  & $1.47\pm0.12\pm0.14$ \\
$\eta^{\prime}\pi^{\pm}$ & $733\pm33$   & 5.4   & $4.02\pm0.27\pm0.30$ \\
\botrule
\end{tabular}}
\end{table}

CLEO has also measured the inclusive $\eta$, $\eta^{\prime}$ and $\phi$ BF's
in $D^0$, $D^+$ and $D_s$ decays\cite{d_incl}. The analysis uses 
281~$\ipb$ of $\psi(3770)$ data and about 200~$\ipb$ recorded at $E_{\rm cm}=4170$~MeV.
The approach is to reconstruct one $D$ meson (called the {\it tag}), and use 
the remaining charged particles and showers to search for the decays:
$\eta\to\gamma\gamma$, $\eta^{\prime}\to\eta\pi^+\pi^-$ and $\phi\to K^+K^-$.
Yields are determined by fitting the invariant mass spectra (for $\eta^{\prime}$, we use the
mass difference, $M_{\eta\pi^+\pi^-}-M_{\eta}$), and performing a sideband subtraction
for the tag-side background. Yields and efficiencies are determined in 2 bins of momentum
for $\eta$ and 5 momentum bins for $\eta^{\prime}$ and $\phi$,
and the resulting partial BF's are then summed to obtain
the total BF. The results are summarize in Table~\ref{tab:dinc}.
The inclusive $\eta$ results include feed-down from $\eta^{\prime}$. Prior
to these measurements, only weak limits or measurements with large uncertainties
existed. These measurements support the expectation that the $s\bar{s}$
component of $\eta$, $\eta^{\prime}$, and $\phi$ mesons lead to a larger production
rate in $D_s$ decays.

\begin{table*}
\tbl{Inclusive branching fractions for $D^0$, $D^+$ and $D_s$ to $\eta$,
$\eta^{\prime}$, and $\phi$ mesons (in percent).
\label{tab:dinc}}
{\begin{tabular}{lccc} 
\toprule
       & $\eta X$ & $\eta^{\prime} X$ & $\phi X$ \\
\hline
$D^0$  & $9.5\pm0.4\pm0.8$  & $2.48\pm0.17\pm0.21$   & $1.05\pm0.08\pm0.07$ \\
$D^+$  & $6.3\pm0.5\pm0.5$  & $1.04\pm0.16\pm0.09$   & $1.03\pm0.10\pm0.07$ \\
$D_s$  & $23.5\pm3.1\pm2.0$ & $8.7\pm1.9\pm0.8$      & $16.1\pm1.2\pm1.1$ \\
\botrule
\end{tabular}}
\end{table*}

\section{Dalitz Analyses}

Multi-body decays are often dominated by, or have significant contributions from
one or more quasi-twobody decays. Dalitz analyses allow for extraction of these
contributing amplitudes and relative strong phases by constructing a total amplitude
which is a coherent sum of complex amplitudes\cite{dalitz_pdg}.  

We first report on a CLEO Dalitz analysis of the decay 
$D^+\to\pi^+\pi^-\pi^+$\cite{d3pi_dalitz}. This
work is fueled by a number of experimental results\cite{e791,besii} that 
support the existence of a low mass scalar state, referred to
as the $\sigma$. FOCUS, which employs a $K$ matrix formalism\cite{focus}, does not
require the inclusion of the $\sigma$ to describe the low $\pi\pi$ mass region
in the $D\to\pi^+\pi^-\pi^+$ decay. 

CLEO-c's $D^+\to\pi^+\pi^-\pi^+$ sample is comprised of about
2600 $D^+\to\pi^+\pi^-\pi^+$ decays and about 2200 background events
after vetoing events consistent with $K^0_S\to\pi^+\pi^-$. Several $\pi^+\pi^-$ resonant
contributions are included in the fit, including $\rho(770)$, $\rho(1450)$, $f_0(980)$,
$f_2(1270)$, $f_0(1370)$, $f_0(1500)$, and $\sigma$. The $\sigma$ is modeled as
a pole in the complex plane\cite{oller} and the $f_0(980)$ using the Flatt\'{e}
parameterization. The resonance parameters are taken from previous 
experiments\cite{pdg04,bes-pars}. The resulting fit fractions and relative
phases are shown in Table~\ref{tab:pi3}. Our data are in general agreement
with those of E791\cite{e791}. In both cases, neglecting this low mass 
contribution gives a poor description of the data. Alternate descriptions
of the low mass $\pi\pi$ region are being investigated.

\begin{table}
\tbl{Preliminary results on the Dalitz analysis of the $D^+\to\pi^+\pi^-\pi^+$ 
decay showing the fit fraction (FF) and phase relative to $\rho(770)\pi^+$. 
CLEO limited are at the 90\% confidence level. We also show the corresponding
fit fractions from E791.
\label{tab:pi3}}
{\begin{tabular}{@{}lccc@{}} 
\toprule
Mode                 &  CLEO FF     & Phase($^o$) &  E791 FF \\
\hline
$\rho(770)\pi^+$     & $20.0\pm2.5$ & $0$(fixed)  & $33.6\pm3.9$ \\
$\sigma\pi^+$        & $41.8\pm2.9$ & $-3\pm4$    & $46.3\pm9.1$ \\
$f_2(1270)\pi^+$     & $18.2\pm2.7$ & $-123\pm6$  & $ 19.4\pm2.5$ \\
$f_0(980)\pi^+$      & $4.1\pm0.9$  & $12\pm11$   & $6.2\pm1.4$ \\
$f_0(1500)\pi^+$     & $3.4\pm1.3$  & $-44\pm21$  &    - \\
Non-resonant         & $<3.5$       &       -     & $7.8\pm6.6$ \\
$\rho(1450)\pi^+$    & $<2.4$       &       -     & $0.7\pm0.8$ \\
\botrule
\end{tabular}}
\end{table}

    CLEO has also performed a Dalitz analysis of the $D^0\to K^+K^-\pi^0$ 
decay\cite{kkpi_dalitz} using $\sim$9~$\ifb$ of data on/just below the $\Upsilon$(4S) .
This final can be used to extract the CP-violating angle,
$\gamma$, through the interference of $B^+\to D^0 K^+$ and $B^+\to\bar{D^0}K^+$\cite{gls}.
We reconstruct and tag the flavor of $D^0$ mesons via the sequential decay $D^{*+}\to D^0\pi^+$.
A sample of 735 events are selected, of which about 130 are background.
The signal is modeled with three Breit-Wigner contributions, 
$D^0\to\phi\pi^0$, $K^{*+}K^-$, and $K^{*-}K^+$, along with a non-resonant (NR)
interfering component. The background shape is determined by a fit to the 
Dalitz plot obtained from $D^0$ candidates in the $\Delta M=M(D^{*+})-M(D^0)$ 
sideband region. The amplitudes, phases and fit fractions are shown in
Table~\ref{tab:kkpi}. For $K^{*-}K^+$, which is most relevant for this analysis,
the uncertainties include both statistical and systematic sources; the others
are only statistical. The amplitude ratio and relative phase between
$K^{*-}K^+$ and $K^{*+}K^-$ are thus found to be:
$|A_{K^{*-}K^+}/A_{K^{*+}K^-}|=0.52\pm0.05\pm0.04$ and
$\delta_{K^{*-}K^+-K^{*+}K^-}=(332\pm8\pm11)^o$. This is the first measurement
of this phase shift, and the amplitude ratio measurement is a significant improvement 
over previous estimates using $D^0\to K^*K$ branching fractions.

We gratefully acknowledge the effort of the CESR staff 
in providing us with excellent luminosity and running conditions,
and the National Science Foundation for support of this work.

\begin{table}
\tbl{Dalitz results of the $D^+\to K^+K^-\pi^0$ decay showing the 
amplitudes, phases and fit fractions (FF) relative to $K^{*+}K^-$.
\label{tab:kkpi}}
{\begin{tabular}{@{}lccc@{}} 
\toprule
Mode           & Amplitude            & Phase($^o$) &        FF \\
\hline
$K^{*+}K^-$     & $1.0$                & $0$(fixed)      & $46.1\pm3.1$ \\
$K^{*-}K^+$    & $0.52\pm0.06$        & $332\pm14$      & $12.3\pm2.2$ \\
$\phi\pi^0$    & $0.64\pm0.04$        & $326\pm9$       & $1.94\pm1.6$ \\
NR   & $5.62\pm0.45$        & $220\pm5$       & $36.0\pm3.7$ \\
\botrule
\end{tabular}}
\end{table}

\end{document}